\documentstyle[fleqn,twoside,amsfonts]{article}

\makeatletter

\newcommand{\bls}[1]{\renewcommand{\baselinestretch}{#1}}

\def\noi{\noindent}

\renewcommand{\section}{\@startsection{section}{1}{0pt}%
        {-3.5ex plus -1ex minus -.2ex}{2.3ex plus .2ex}%
        {\large\bf\protect\raggedright}}

\renewcommand{\subsection}{\@startsection{subsection}{2}{0pt}%
        {-3ex plus -1ex minus -.2ex}{1.4ex plus .2ex}%
        {\normalsize\bf\protect\raggedright}}

\renewcommand{\thesubsubsection}%
        {\arabic{section}.\arabic{subsection}.\arabic{subsubsection}.}

\renewcommand{\@oddhead}{\raisebox{0pt}[\headheight][0pt]{%
   \vbox{\hbox to\textwidth{\rightmark \hfil \rm \thepage \strut}\hrule}}}
\renewcommand{\@evenhead}{\raisebox{0pt}[\headheight][0pt]{%
   \vbox{\hbox to\textwidth{\thepage \hfil \leftmark \strut}\hrule}}}
\newcommand{\heads}[2]{\markboth{\protect\small\it #1}{\protect\small\it
#2}}
\newcommand{\Acknow}[1]{\subsection*{Acknowledgement} #1}

\newcommand{\Title}[1]{\noi {\Large #1} \\}

\newcommand{\Authors}[4]{\noi
        {\large\bf #1\dag\ #2\ddag}\medskip\begin{description}
        \item[\dag]{\it #3} \item[\ddag]{\it #4}\end{description}}

\newcommand{\Abstract}[1]{\vskip 2mm \begin{center}
        \parbox{16.4cm}{\small\noi #1} \end{center}\medskip}

\newcommand{\foom}[1]{\protect\footnotemark[#1]}
\newcommand{\foox}[2]{\footnotetext[#1]{#2}}
\newcommand{\email}[2]{\footnotetext[#1]{e-mail: #2}}
\newcommand{\eps}{ \varepsilon }

\newcommand{\Theorem}[2]{\medskip\noi {\bf #1. \ }{\sl #2}\medskip}

\newcommand{\sect}[1]{Sec.\,#1}
\def\nq{\hspace*{-1em}}
\def\nqq{\hspace*{-2em}}
\def\nhq{\hspace*{-0.5em}}

\def\cm{\hspace*{1cm}}

\def\al{&\nhq}
\def\lal{&&\nqq {}}

\def\eqs{Eqs.\,}
\def\beq{\begin{equation}}
\def\eeq{\end{equation}}
\def\bear{\begin{eqnarray}}
\def\bearr{\begin{eqnarray} \lal}
\def\ear{\end{eqnarray}}
\def\earn{\nonumber \end{eqnarray}}

\def\nnn{\nonumber\\ \lal }

\def\yy{\\[5pt] {}}

\def\eql{\al =\al}

\def\e{{\,\rm e}}

\def\sign{\mathop{\rm sign}\nolimits}

\def\dim{\mathop{\rm dim}\nolimits}

\makeatother

\def\beq#1{\begin{equation}\label{#1}}
\def\eeq{\end{equation}}
\def\ber#1{\begin{eqnarray}\label{#1} \nqq}
\def\eer{\end{eqnarray}}

\catcode`\@=11 \@addtoreset{equation}{section}\catcode`\@=12

\newcommand{\N}{{\mathbb N}}
\newcommand{\R}{{\mathbb R}}
\newcommand{\C}{{\mathbb C}}

\newcommand{\p}{\partial}

\newcommand{\tri}{\Delta}

\def\brane{$p$-brane}

\heads{V.D.Ivashchuk, V.S. Manko and V.N.Melnikov}
       {PN parameters for general black hole and
       spherically symmetric $p$-brane solutions}

\bls{0.98}
\begin{document}
\twocolumn[

\thispagestyle{empty}

\Title{POST-NEWTONIAN PARAMETERS FOR GENERAL BLACK HOLE \yy
AND SPHERICALLY SYMMETRIC $p$-BRANE SOLUTIONS}

\Authors{V.D. Ivashchuk,\foom 1} {V.S. Manko\foom 2 and V.N. Melnikov\foom
3}
{Centre for Gravitation and Fundamental Metrology,
VNIIMS, 3-1 M. Ulyanovoy Str.,
Moscow 117313, Russia \\ and
Institute of Gravitation and Cosmology,
Peoples' Friendship University of Russia,\\
6 Miklukho-Maklaya St., Moscow 117198, Russia}
{Depto. de Fisica, CINVESTAV-IPN, Apartado Postal 14-740,
Mexico 07000, D.F.}

\Abstract
{Black hole $p$-brane solutions for a wide class of intersection rules are
considered. The solutions are defined on a manifold which contains a
product of $(n-1)$ Ricci-flat ``internal'' spaces. The post-Newtonian
parameters $\beta$ and $\gamma$ corresponding to a 4-dimensional section
of the metric for general intersection rules are studied. It is shown
that  $\beta$ does not depend but $\gamma$ depends on \brane\
intersections.  For ``block-orthogonal" intersection rules
spherically symmetric solutions are considered, and explicit relations for
post-Newtonian parameters are obtained. The bounds on parameters of
solutions following from observational restrictions in the Solar system are
presented.}

] 

\email 1 {ivas@rgs.phys.msu.su}
\email 2 {vsmanko@fis.cinvestav.mx}
\foox 3
{Permanent address: Centre for Gravitation and Fundamental Metrology,
VNIIMS, 3-1 M. Ulyanovoy Str., Moscow 117313, Russia  and
Institute of Gravitation and Cosmology, PFUR,
6 Miklukho-Maklaya St., Moscow 117198, Russia}

\section{Introduction}

Multidimensional gravitational models
(see, e.g., [1--5] and references therein)
are rather popular at present.
There exist two main approaches within the multidimensional
paradigm: ({\bf i}) extra dimensions are considered as real space-time
dimensions; ({\bf ii}) extra dimensions are considered as some imaginary
(e.g. mirror, or virtual) reality, as a mathematical tool
for generating the 4-metric and extra (dilatonic) scalar fields.
In the second approach the 4-dimensional metric is a ``real physical
object" that should be extracted in some way from the multidimensional one.

In both cases we deal with exact solutions to multidimensional field
equations with 4-dimensional sections which may describe some astrophysical
objects (e.g. the Sun, neutron stars, etc.) and the motion of test
bodies (e.g. Mercury). The 4-metric is usually considered to be a small
``deformation" of the original  Schwarzschild one and is governed by some
parameters related to the post-Newtonian (PN) ones (for definition see
\sect 2).  Applying the classical tests of general relativity:
gravitational redshift, light deflection, perihelion advance and time
delay, (see \cite{Wil,Dam}) along with the geodetic precession test, we can
find the restrictions on the PN parameters.

In \cite{BIMt,M} the PN parameters and classical tests were considered
(in the framework of the approach ({\bf ii})) for a metric generalizing the
Schwarzschild solution to the case of several Ricci-flat internal spaces.
Later on, Kramer's more special 5-dimensional solution (also known as the
Gross-Perry-Sorkin-Davidson-Owen one) was considered in [9--11]
for the approach ({\bf i}) and the classical tests  (e.g. perihelion
advance and geodetic precession) were generalized to situations in which
the components of momentum and spin along the extra coordinate do not
vanish.

In this paper we deal with 4-dimensional sections of mutidimensional
spherically symmetric and black hole solutions with generalized $p$-branes
charged by fields of forms. In \sect 2 black hole  $p$-brane  solutions
for general intersection rules are considered and the PN parameters are
calculated.  In \sect 3 more general spherically symmetric solutions
are considered, but for more special ``block-orthogonal" intersection
rules, and explicit relations for PN parameters are obtained.
In both sections the bounds on the solution gravitation parameter following
>from observational restrictions in the Solar system are considered.

\section{Black hole p-brane solutions}

\subsection{Solutions}

We consider a model governed by the action
\bearr\label{i.1}
S=\int d^Dx \sqrt{|g|}\biggl\{R[g]-h_{\alpha\beta}g^{MN}\p_M\varphi^\alpha
 \p_N\varphi^\beta
\nnn \cm
-\sum_{a\in\tri}\frac{\theta_a}{n_a!}
  \exp[2\lambda_a(\varphi)](F^a)^2\biggr\}
\ear
where $g=g_{MN}(x)dx^M\otimes dx^N$ is a D-dimensional metric,
$\varphi=(\varphi^\alpha)\in\R^l$ is a vector of scalar (dilatonic) fields,
$(h_{\alpha\beta})$ is constant symmetric
non-degenerate $l\times l$ matrix $(l\in \N)$, $\theta_a=\pm1$,
$F^a = dA^a
=  \frac{1}{n_a!} F^a_{M_1 \ldots M_{n_a}}
dz^{M_1} \wedge \ldots \wedge dz^{M_{n_a}}$
is an $n_a$-form ($n_a\ge1$), $\lambda_a$ is a
1-form on $\R^l$: $\lambda_a(\varphi)=\lambda_{\alpha a}\varphi^\alpha$,
$\alpha=1,\dots,l$; we denote $|g| =   |\det (g_{MN})|$ and $(F^a)^2_g  =
F^a_{M_1 \ldots M_{n_a}} F^a_{N_1 \ldots N_{n_a}}
g^{M_1 N_1} \cdots g^{M_{n_a} N_{n_a}}$,
$a \in \tri$. Here $\tri$ is some finite set.

Let us consider a family of black-hole
solutions to the field equations corresponding to the action
(\ref{i.1}). These solutions are defined on the manifold
\bearr\label{i.2}
M =    (R_0, + \infty)  \times
(M_1 = S^{d_1}) \times (M_2 = \R) \times  \ldots \times M_{n},\nnn
\ear
and have the following form:
\bear\label{i.3}
 g \eql \Bigl(\prod_{s \in S} H_s^{2 h_s d(I_s)/(D-2)} \Bigr)
  \biggl\{ F^{-1} dR \otimes dR
\nnn \qquad
 + R^2  d \Omega^2_{d_1}
 -  \Bigl(\prod_{s \in S} H_s^{-2 h_s} \Bigr) F  dt \otimes dt
\nnn \cm\quad
 + \sum_{i = 3}^{n} \Bigl(\prod_{s\in S}
   H_s^{-2 h_s \delta_{iI_s}} \Bigr) g^i  \biggr\},
\\  \label{i.31}
    \e^{\varphi^\alpha} \eql
 \prod_{s\in S} H_s^{h_s \chi_s \lambda_{a_s}^\alpha},
\\  \label{i.32a}
F^a  \eql  \sum_{s \in S} \delta^a_{a_s} {\cal F}^{s},
\ear
where
\bear\label{i.9}
{\cal F}^s \eql \frac{Q_s}{R^{d_1}}
\Bigl( \prod_{s' \in S}  H_{s'}^{- A_{s s'}} \Bigr) dR \wedge\tau(I_s),
  \ \ s \in S_e, \\
\label{i.10}
{\cal F}^s \eql  Q_s \tau(\bar I_s), \cm s \in S_m.
\ear
Here
$Q_s \neq 0$ ($s \in S$)
are charges, $R_0 > 0$, $R_0^{\bar d} = 2 \mu > 0$,
$\bar d = d_1 -1$, $F = 1 - 2\mu/ R^{\bar d}$.

In  (\ref{i.3}),
$g^i = g^i_{m_i n_i}(y^i)$ is a Ricci-flat  metric on $M_{i}$,
$i=  2,\ldots,n$,  and
\beq{1.11}
\delta_{iI}=  \sum_{j\in I} \delta_{ij}
\eeq
is the indicator of $i$ belonging
to $I$: $\delta_{iI}=  1$ for $i\in I$ and $\delta_{iI}=  0$ otherwise.
We note that here all $p$-branes do not ``live'' in  $M_1$.

The  $p$-brane  set  $S$ is by definition
\beq{i.6}
 S=  S_e \cup S_m, \quad
 S_v=  \cup_{a\in\tri}\{a\}\times\{v\}\times\Omega_{a,v},
\eeq
$v=  e,m$ and $\Omega_{a,e}, \Omega_{a,m} \subset \Omega$,
where $\Omega =   \Omega(n)$  is the set of all non-empty
subsets of $\{ 2, \ldots,n \}$. Thus, any $s \in S$
has the form $s =   (a_s,v_s, I_s)$,
where $a_s \in \tri$, $v_s =  e,m$ and $I_s \in \Omega_{a_s,v_s}$.
The sets $S_e$ and $S_m$ define electric and magnetic $p$-branes,
respectively. In  (\ref{i.31}) $\chi_s  = +1, -1$ for $s \in S_e, S_m$
respectively.

All the  manifolds $M_{i}$, $i > 1$ are assumed to be oriented and
connected and  the volume $d_i$-forms
\beq{i.12}
\tau_i  \equiv \sqrt{|g^i(y_i)|}
\ dy_i^{1} \wedge \ldots \wedge dy_i^{d_i},
\eeq
are well defined for all $i=  1,\ldots,n$.
Here $d_{i} =   \dim M_{i}$,
$M_1 = S^{d_1}$, $d_1 > 1$;
$D =   1 + \sum_{i =   1}^{n} d_{i}$, and for any
$I =   \{ i_1, \ldots, i_k \} \in \Omega$, $i_1 < \ldots < i_k$,
we denote
\bearr\label{i.13}
  \tau(I) \equiv \tau_{i_1}  \wedge \ldots \wedge \tau_{i_k},
  \qquad d(I) \equiv   \sum_{i \in I} d_i.
\ear
In (\ref{i.10}) $\bar I\equiv\{1,\ldots,n\}\setminus I$.

The parameters  $h_s$ appearing in the solution
satisfy the relations
\beq{i.16}
  h_s = K_s^{-1}, \qquad  K_s = B_{s s},
\eeq
where
\bearr\label{i.17} \nq
 B_{ss'} \equiv
 d(I_s\cap I_{s'})+\frac{d(I_s)d(I_{s'})}{2-D}
  {+}\chi_s\chi_{s'}\lambda_{\alpha a_s}\lambda_{\beta a_{s'}}
  h^{\alpha\beta},     \nnn
\ear
$s, s' \in S$, with $(h^{\alpha\beta})=(h_{\alpha\beta})^{-1}$.
Here we assume that
\beq{i.17a}
 ({\bf i}) \qquad B_{ss} \neq 0,
\eeq
for all $s \in S$, and
\beq{i.18b}
({\bf ii}) \qquad {\rm det}(B_{s s'}) \neq 0,
\eeq
i.e., the matrix $(B_{ss'})$ is non-degenerate. Let
\beq{i.18}
(A_{ss'}) = \left( 2 B_{s s'}/B_{s' s'} \right).
\eeq
Here  some ordering in $S$ is assumed.

The functions $H_s = H_s(z) > 0$, $z = 2\mu/ R^{\bar d} \in (0,1) $ obey
the
equations
\beq{i3.1}
  \frac{d}{dz} \left( \frac{(1 -z)}{H_s}
  \frac{d}{dz} H_s \right) = B_s
\prod_{s' \in S}  H_{s'}^{- A_{s s'}},
\eeq
equipped with the boundary conditions
\bearr\label{i3.2a}
H_{s}(1 -0) = H_{s0} \in (0, + \infty), \\ \label{i3.2b}\lal
H_{s}(+0) = 1, \\ \label{i1.21}  \lal
  B_s = K_s \eps_s Q_s^2 /(2 \mu)^2,
\ear
$s \in S$. Here
\beq{i1.22}
\eps_s=(-\eps[g])^{(1-\chi_s)/2}\eps(I_s) \theta_{a_s},
\eeq
$s\in S$, $\eps[g]\equiv\sign\det(g_{MN})$. More explicitly,
(\ref{i1.22}) reads: $\eps_s=\eps(I_s) \theta_{a_s}$ for
$v_s = e$ and $\eps_s=-\eps[g] \eps(I_s) \theta_{a_s}$, for
$v_s = m$.

\eqs (\ref{i3.1}) are equivalent to special Toda-type equations.
The first boundary condition (\ref{i3.2a}) guarantees
the existence of a horizon at  $R^{\bar{d}} = 2 \mu$
for the metric (\ref{i.3}).
The second condition (\ref{i3.2b}) ensures the asymptotic (at $R \to
+\infty$) flatness of the $(2 + d_1)$-section of the metric.

Due to (\ref{i.9}) and  (\ref{i.10}), the
$p$-brane worldsheet dimension $d(I_s)$ is defined by
\beq{i.16a}
d(I_s)=  n_{a_s}-1, \quad d(I_s)=   D- n_{a_s} -1,
\eeq
for $s \in S_e, S_m$ respectively.
For a $p$-brane: $p =   p_s =   d(I_s)-1$.

The solutions are valid if the following  restrictions on the sets
$\Omega_{a,v}$ are imposed.
(These restrictions guarantees the block-diagonal structure
of the stress-energy tensor, as for the metric, and the existence of
a $\sigma$-model representation \cite{IMC}).
We denote $w_1\equiv\{i|i\in\{1,\dots,n\},\quad d_i=1\}$, and
$n_1=|w_1|$ (i.e. $n_1$ is the number of 1-dimensional spaces among
$M_i$, $i=1,\dots,n$).

{\bf Restriction .} Let 1a) $n_1\le1$ or 1b) $n_1\ge2$ and for
any $a\in\tri$, $v\in\{e,m\}$, $i,j\in w_1$, $i<j$, there are no
$I,J\in\Omega_{a,v}$ such that $i\in I$, $j\in J$ and $I\setminus\{i\}=
J\setminus\{j\}$.

This restriction is  satisfied in the non-composite case: $|\Omega_{a,v}| =
1$, (i.e. when there is only one $p$-brane for each value of the colour
index $a$, $a\in\tri$); The restriction forbids certain intersections of
two $p$-branes with the same colour index for  $n_1 \geq 2$.

The Hawking ``temperature" corresponding to the solution is found to be
\beq{i2.36}
T_H=   \frac{\bar{d}}{4 \pi (2 \mu)^{1/\bar{d}}}
\prod_{s \in S} H_{s0}^{- h_s},
\eeq
where $H_{s0}$ are defined in (\ref{i3.2a})

\medskip\noi
{\bf ``Block-orthogonal" black holes \cite{CIM}.} Let
\beq{5.3an}
S=S_1\sqcup\dots\sqcup S_k,
\eeq
$S_i\ne\emptyset$, $i=1,\dots,k$, and
\beq{5.4n}
{\bf (iii)} \
(U^s,U^{s'})= 0
\eeq
for all $s \in S_i$,
$s' \in S_j$,
$i\ne j$; $i,j=1,\dots,k$.
In \cite{Br,IMJ2} ``block-orthogonal" solutions were obtained:
\beq{i3.11a}
 H_{s}(z) = (1 + \bar P_s z)^{b_0^s},
\eeq
$s \in S$, where
\beq{i3.11b}
b_0^s = 2 \sum_{s' \in S} A^{s s'},
\eeq
($(A^{s s'}) = (A_{s s'})^{-1}$) and
$\bar P_s = \bar P_{s'}$, $s,s' \in S_i$, $i = 1, \ldots, k$.

Let
$(A_{s s'})$ be  a Cartan matrix  for a  finite-dimen\-si\-onal
semisimple Lie  algebra $\cal G$. In this case all powers in
(\ref{i3.11a}) are positive integers \cite{GrI}:
\beq{i3.11}
b_0^s = n_s \in \N,
\eeq
and hence all functions $H_s$ are polynomials, $s \in S$.

\medskip\noi
{\bf Conjecture.} {\em Let
$(A_{s s'})$ be  a Cartan matrix  for a  semisimple
finite-dimensional Lie algebra $\cal G$. Then  the solutions
of \eqs (\ref{i3.1})--(\ref{i3.2b})  are polynomials
\beq{i3.12}
H_{s}(z) = 1 + \sum_{k = 1}^{n_s} P_s^{(k)} z^k,
\eeq
where $P_s^{(k)}$ are constants,
$k = 1,\ldots, n_s$, $P_s^{(n_s)} \neq 0$,  $s \in S$ }.

In \cite{IMPol}  this Conjecture was verified for simple Lie algebras
$A_{m}= sl(m+1, \C)$, $m \geq 1$.  In extremal case ($\mu = + 0$) an a
analogue of this conjecture was suggested previously in \cite{LMMP}.

\subsection{Post-Newtonian approximation}

Let $d_1 =   2$. Here we consider the 4-dimensional section of
the metric (\ref{i.3}), namely,
\bearr\label{i5.1}\nhq
g^{(4)} =   U \biggl\{ \frac{dR
\otimes dR}{1 - 2\mu / R} + R^2  d \Omega^2_{2}
- U_1\! \Bigl(1 - \frac{2\mu}{R} \Bigr)\!  dt \otimes dt \biggr\}
\nnn
\ear
in the range $R > 2\mu$, where
\bear\label{i5.1a}
 U \eql   \prod_{s \in S} H_s^{2 d(I_s) h_s/(D-2)},
   \\  \label{i5.1b}
 U_1 \eql \prod_{s \in S} H_s^{-2 h_s}.
\ear

Let us  imagine that some real astrophysical objects (e.g. stars) may
be described (or approximated) by the 4-dimensional  physical  metric
(\ref{i5.1}), i.e. they are traces of extended multidimensional objects
(charged $p$-brane black holes).

In the post-Newtonian approximation
we restrict ourselves to the first two powers of $1/R$, i. e.
\beq{i5.2a}
H_s = 1 + \frac{P_s}{R}  + \frac{P_s^{(2)}}{R^2} + o(\frac{1}{R^3}),
\eeq
for $R \to + \infty$, $s \in S$.

Introducing a new radial variable $\rho$ by the relation
\beq{i5.2}
R =   \rho \left(1 + \frac{\mu}{2\rho}\right)^2,
\eeq
($\rho > \mu/2$), we may rewrite the metric (\ref{i5.1}) in a 3-dimensional
conformally flat form and  calculate the post-Newtonian parameters $\beta$
and $\gamma$ (Eddington parameters) using the following standard relations:
\bear\label{i5.4}
 g^{(4)}_{00} \eql - (1 -  2 V + 2 \beta V^2 ) + O(V^3),
\\                       \label{i5.5}
 g^{(4)}_{ij} \eql  \delta_{ij}(1 + 2 \gamma V ) + O(V^2),
\ear
$i,j =   1,2,3$, where $V =   GM/\rho$ is  Newton's potential, $G$ is the
gravitational constant and $M$ is the gravitational mass. From
(\ref{i5.4})-(\ref{i5.5}) we deduce the formulae
\beq{i5.7}
  GM =   \mu + \sum_{s \in S} h_s P_s
  \left(1 -  \frac{d(I_s)}{D-2} \right)
\eeq
and
\bear\label{i5.8}
 \beta - 1 \eql
  \frac{1}{2(GM)^2} \sum_{s \in S}  h_s (P_s^2 + 2 \mu P_s
\nnn\cm\cm
 - 2 P_s^{(2)}) \Bigl(1 -  \frac{d(I_s)}{D{-}2} \Bigr), \\
\label{i5.9}
 \gamma - 1 \eql - \frac{1}{GM} \sum_{s \in S} h_s P_s \left(1 -  2
  \frac{d(I_s)}{D-2} \right).
\ear

For special ``block-orthogonal'' solutions this relation coincides with
that of \cite{CIM}. For a more general spherically symmetric case it was
obtained in \cite{IMJ2} (see next section).

The parameter $\beta$ is determined by the squared charges
$Q_s^{2}$ of the $p$-branes (or, more correctly, by the charge densities)
and the signature parameters $\eps_s$.

Now, we will show that in the general case, as in \cite{CIM}, the
following theorem is valid:

\Theorem{Theorem} {The parameter $\beta$ does not depend on
the dimensions of $p$-brane intersections}.

\noi
{\bf Proof.}
>From (\ref{i3.1}) we get in the zero order of a $z$-decomposition:
\bear\label{i5.10}
 P_s^2 +2 \mu P_s - 2 P_s^{(2)} = - K_s \eps_s Q_s^2,
\ear
$s \in S$.  Hence,
\bear\label{i5.11}
\beta - 1 =   \frac{1}{2(GM)^2} \sum_{s \in S} (- \eps_s) Q_s^{2}
\left(1 -  \frac{d(I_s)}{D-2} \right).
\ear
Thus $\beta$ depends on  $Q_s/GM$, i.e., on the ratios of the charges and
the mass. It does not depend on the intersections.  The theorem is proved.

\medskip
The parameter $\beta$ is obtained without knowledge
of the general solution for $H_s$ and does not depend
on the quasi-Cartan matrix and hence on \brane\ intersections.
The parameter $\gamma$ depends on the ratios $P_s/GM$, where $P_s$
are functions of $GM$, $Q_s$ and also $A = (A_{s s'})$.
Thus it depends on the intersections. The calculation of $\gamma$ needs an
exact solution for the radial functions $H_s$.

For the most physically interesting $p$-brane solutions, $\eps_s = -1$ and
$d(I_s) < D - 2$ for all $s \in S$, whichimplies
\beq{i5.12}
 \beta > 1.
\eeq

Using the relations (\ref{i5.9}) and (\ref{i5.11}), we can obtain bounds on
$\beta$ and $\gamma$ which may be compared with the experimental data.

\medskip\noi
{\bf Observational restrictions.} The observations in the Solar system
give tight constraints on the Eddington parameters \cite{Dam}:
\bear\label{i.14}
 \gamma \eql 1.000 \pm 0.002 \\
\label{i.15}
 \beta \eql 0.9998 \pm 0.0006.
\ear
The first restriction is a result of the Viking time-delay experiment
\cite{Re}. The second one follows from (\ref{i.14}) and the analysis of the
lunar laser ranging data. In this case a high precision test
based on the calculation of the combination $(4\beta - \gamma - 3)$,
appearing in the Nordtvedt effect \cite{N}, is used \cite{Di}.

For small enough $p_s = P_s/GM$ and $P_s^{(2)} \sim P_s^{2}$
we get $GM \sim \mu$ and hence
\bear\label{i.17c}
 \beta - 1 \lal \sim  \sum_{s \in S} h_s p_s
  \left(1 -   \frac{d(I_s)}{D-2} \right),
\\ \label{i.17d}
 \gamma - 1 \lal \sim - \sum_{s \in S} h_s p_s
  \left(1 -  2 \frac{d(I_s)}{D-2} \right) ,
\ear
i.e., $\beta -1$ and $\gamma - 1$ are of the same order.
Thus, for small enough $p_s $, it is possible to fit
the Solar-system restrictions (\ref{i.14}) and (\ref{i.15}).

\section{Spherically symmetric solutions in the block-orthogonal case}

\subsection{Solutions}

Here we consider the spherically symmetric solutions
in the block-orthogonal case (\ref{5.3an})--(\ref{5.4n}).

The solutions for the metric and scalar fields
may be written as follows:
\bear\label{5.72n}
g \eql \Bigl(\prod_{s \in S}
 \bar{H}_s^{2 \eta_s d(I_s)\nu_s^2/(D-2)} \Bigr)
  \biggl\{ F^{b^1 -1} dR \otimes dR
\nnn\quad
  + R^2  F^{b^1} d \Omega^2_{d_1}
    - \Bigl(\prod_{s \in S} \bar{H}_s^{-2 \eta_s \nu_s^2} \Bigr)
   F^{b^1}  dt \otimes dt
\nnn\quad
 + \sum_{i = 3}^{n} \Bigl(\prod_{s\in S}
    \bar{H}_s^{-2 \eta_s \nu_s^2 \delta_{iI_s}} \Bigr)
     F^{b^i} g^i  \biggr\},
\\  \label{5.73}
\varphi^\alpha \eql
 \sum_{s\in S} \eta_s \nu_s^2 \chi_s \lambda_{a_s}^\alpha
  \ln \bar{H}_s + \frac{1}{2} b^{\alpha} \ln F,
\ear
where $F = 1 - 2\mu/R^{\bar d}$,
\bear \label{5.73b}
 \bar{H}_s \eql \hat{H}_s  F^{(1-b_s)/2},
\\ \label{5.74}
 \hat{H}_s \eql 1 +  \hat{P}_s \frac{(1 - F^{b_s})}{2 \mu b_s },
\ear
$s \in S$.
Here
\beq{5.42p}
 \sum_{s' \in S} (U^s, U^{s'}) \eta_{s'} \nu_{s'}^2 = 1
\eeq
for all $s \in S$. The parameters $b_s$ and $\hat{P}_s$
coincide within the blocks:
\beq{5.42t}
 b_s = b_{s'},  \qquad \hat{P}_s = \hat{P}_{s'},
\eeq
$s,s' \in S_i$, $i = 1, \ldots, k$.
The parameters $b_s, b^i, b^{\alpha}$ obey the relations
\bear\label{5.68b}
 U^1(b) \eql  -b^1+ \sum_{j=1}^n d_j b^j= 1,
\\ \label{5.68c}
 U^s(b) \eql
 \sum_{i\in I_s}d_i b^i-\chi_s \lambda_{a_s \alpha} b^\alpha= 1,
\ear
$s\in S$, and
\bearr\label{5.68d}
 \sum_{s \in S} \eta_s \nu_s^2 (b_s^2 - 1) +
 h_{\alpha\beta} b^\alpha b^\beta +
   \sum_{i=2}^n d_i (b^i)^2
\nnn         \cm
+ \frac1{d_1-1} \left(\sum_{i=2}^n d_i b^i \right)^2 = \frac{d_1}{d_1-1}.
\ear
The fields of forms in (\ref{i.32a})  are given by the relations
\bear\label{5.75a}
 {\cal F}^{s}\eql d\Phi^{s}\wedge\tau(I_s), \\ \label{5.75b}
 {\cal F}^{s}\eql \exp[- 2 \lambda_{a_s}(\varphi)]
  *\left(d\Phi^s \wedge \tau(I_s) \right),
\ear
where $* = *[g]$ is the Hodge operator and
\bear\label{5.75}
 \Phi^s \eql {\nu_s}/H'_s,
\\ \label{5.76}
 H_s^{'}\eql \Bigl[1 - P_s^{'} \hat{H}_s^{-1}
  \frac{(1 - F^{b_s})}{2 \mu b_s }  \Bigr]^{-1},
\ear
$s \in S$. Here
\beq{5.78}
 (P_s^{'})^2 = - \eps_s \eta_s \hat{P}_s (\hat{P}_s +2 b_s \mu),
\eeq
$s \in S$.

\subsection{Post-Newtonian parameters}

Let $d_1 = 2$. Consider the 4-dimensional section of the
metric (\ref{5.72n})
\bearr\label{4.1}\nq
 g^{(4)} = U \Bigl\{ F^{b^1-1} dR \otimes dR
  {+} F^{b^1} R^2  d \hat{\Omega}^2_{2}
  {-} U_2 F^{b^2} dt {\otimes} dt \Bigr\} \nnn
\ear
where  $F = 1- 2 \mu/R$ and
\bear\label{4.1a}
 U \eql \prod_{s \in S} \bar{H}_s^{2 \eta_s d(I_s) \nu_s^2/(D-2)},
\\ \label{4.2a}
 U_2 \eql \prod_{s \in S} \bar{H}_s^{-2 \eta_s \nu_s^2},
\ear
$R > 2\mu$.

Introducing the radial variable $\rho$
by the relation (\ref{i5.2}), we rewrite the metric (\ref{4.1})
in a 3-dimensional conformally-flat form:
\bearr\label{4.3}
 g^{(4)} = U \Biggl\{ - U_2 F^{b^2} dt \otimes dt
\nnn \cm
 + F^{b^1} \left(1 + \frac{\mu}{2 \rho} \right)^4
   \delta_{ij} dx^i \otimes dx^j \Biggr\},
\\ \label{4.3a} \lal
 \ F =  \left(1 - \frac{\mu}{2 \rho} \right)^2
  \left(1 + \frac{\mu}{2 \rho} \right)^{-2},
\ear
where $\rho^2 =|x|^2 = \delta_{ij}x^i x^j$ ($i,j = 1,2,3$).

From
(\ref{4.3})--(\ref{4.3a}) and (\ref{i5.4})--(\ref{i5.5})
we get for the gravitational mass:
\bearr\label{4.7}
 GM = \mu b^2
\nnn \quad
 + \sum_{s \in S} \eta_s \nu_s^2[ \hat{P}_s + (b_s - 1)\mu ]
   \left(1 -  \frac{d(I_s)}{D-2} \right)
\ear
and, for $GM \neq 0$,
\bearr\label{4.8} \nq
 \beta - 1 = \frac{1}{2(GM)^2} \sum_{s \in S}  \eta_s \nu_s^2
 \hat{P}_s (\hat{P}_s
 {+} 2 b_s \mu) \left(1 -  \frac{d(I_s)}{D{-}2} \right),
\nnn
\\ \label{4.9} \lal \nq
\gamma - 1 = - \frac{1}{GM} \bigg[ \mu (b^1 + b^2 - 1)
\nnn\
 +  \sum_{s \in S} \eta_s \nu_s^2 [\hat{P}_s + (b_s -1)\mu ]
  \left(1 -  2 \frac{d(I_s)}{D-2} \right) \bigg].
\ear

As follows from  (\ref{5.78}), (\ref{4.8})
and the inequalities  $d(I_s) < D - 2$ (for all $s \in S$),
\bear\label{4.10}
 \beta > 1\quad {\rm \ if \  all} \quad \eps_s = -1,
\\ \label{4.10a}
 \beta < 1\quad {\rm \ if \  all} \quad \eps_s = +1.
\ear
There exists a large variety of configurations
with $\beta = 1$ when the relation $\eps_s = {\rm const}$
is broken.

There also exist non-trivial $p$-brane configurations with $\gamma =1$.

\Theorem{Proposition}
{Let the set of $p$-branes consist of
several pairs of electric and magnetic branes. Let
any such pair $(s, \bar{s} \in S)$  correspond to the same colour index,
i.e. $a_s = a_{\bar{s}}$, and $\hat{P}_s = \hat{P}_{\bar{s}}$,
$b_s = b_{\bar{s}}$, $\eta_s \nu_s^2 = \eta_{\bar{s}} \nu_{\bar{s}}^2$.
Then  for $b^1 + b^2 = 1$  we get
\beq{4.13}
 \gamma = 1.
\eeq
}

The Proposition can be readily proved using the relation
$d(I_s) + d(I_{\bar{s}}) = D - 2$, following from (\ref{i.16a}).

For small enough
$\hat{p}_s = \hat{P}_s/GM$, $b_s - 1$, $b^2 - 1$, $b^i$ ($i > 2$)
of the same order we get $GM \sim \mu$ and hence
\bearr\label{4.17}
 \beta - 1 \sim  \sum_{s \in S} \eta_s \nu_s^2 \hat{p}_s
  \left(1 -   \frac{d(I_s)}{D-2} \right)
\\ \label{4.17a} \lal
 \gamma - 1 \sim -b^1 - b^2 + 1
\nnn
 - \sum_{s \in S} \eta_s \nu_s^2 [\hat{p}_s + (b_s -1)]
  \left(1 -  2 \frac{d(I_s)}{D-2} \right) ,
\ear
i.e., $\beta -1$ and $\gamma - 1$ are of the same order. Thus for small
enough $\hat{p}_s $, $b_s - 1$, $b^2 - 1$, $b^i$ ($i > 2$) it is possible
to fit the Solar-system restrictions (\ref{i.14}) and (\ref{i.15}).

There also exists another opportunity to satisfy these restrictions.

\medskip\noi
{\bf One-brane case.}
Consider the special  case of a single $p$-brane.
In this case we have
\beq{4.18}
 \eta_s \nu^{-2}_s =
  d(I_s) \left(1 -  \frac{d(I_s)}{D-2} \right) + \lambda^2.
\eeq
The relations (\ref{4.8}),  (\ref{4.9})  and   (\ref{4.18})
imply that for large enough values of the
dilatonic coupling constant squared $\lambda^2$
and $b^1 + b^2 = 1$ it is possible to ``fine-tune"  the parameters
$(\beta, \gamma)$ near the point (1, 1) even if the parameters  $\hat{P}_s$
are large.

\section{Conclusions}

We have considered a family of black-hole solutions  from \cite{IMPol} with
intersecting $p$-branes with next to arbitrary intersection rules.  The
metric of the solutions  contains  $n -1$ Ricci-flat ``internal'' space
metrics. The solutions are defined up to  a set of functions $H_s$ obeying
a set of equations with certain boundary conditions. In \cite{IMPol}  a
conjecture on the polynomial structure of  $H_s$ for intersections related
to semisimple Lie algebras was suggested and proved for the $A_m$
series.  Here we have presented the post-Newtonian parameters $\beta$ and
$\gamma$ corresponding to the 4-dimensional section of the metric.  The
parameter $\beta$ is written in terms of ratios of the physical parameters:
the charges $Q_s$ and the mass $M$. Thus $\beta$ does not depend on
\brane\ intersections. Unlike that, the parameter $\gamma$ is
intersection-dependent, and its calculation requires an exact solution for
the radial functions $H_s$.  We have considered the post-Newtonian
parameters for spherically symmetric solutions in the block-orthogonal
case and singled out the ``corridors" for the $b$-parameters which follow
>from the Solar-system observational restrictions.

\Acknow
{This work was supported in part by the Russian Ministry of
Science and Technology, Russian Basic Research Foundation
(Grant No. 98-02-16414), Project SEE and CONACYT, Mexico.}

\end{document}